\documentclass[12pt,a4paper]{article}
\usepackage{jheppub}
\usepackage{lscape}
\usepackage{subfigure}
\usepackage{multirow}
\long\def\symbolfootnote[#1]#2{\begingroup%
\def\thefootnote{\fnsymbol{footnote}}\footnote[#1]{#2}\endgroup}

\def\hermesauthor[#1]#2{{#2}$^{\, #1}$}
\def\hermesinstitute[#1]#2{$^{#1\,}$ {#2}\\}

\renewcommand{\thefootnote}{\alph{footnote}}
\def\nowat[#1]#2{\(^,\)\footnote[#1]{#2}}
\def\amp2{{\cal T}}

\DeclareFixedFont{\boldscsixteen}{T1}{phv}{b}{sc}{16pt}
\DeclareFixedFont{\boldscfourteen}{T1}{phv}{b}{sc}{14pt}

\newcommand{\etal}{\emph{et al.}}

\preprint{DESY Report 13-188}

\title{\boldmath Beam-helicity asymmetry in associated electroproduction of real photons
$ep\rightarrow e\gamma \pi N$ in the $\Delta$-resonance region}

\collaboration{The {\boldscsixteen Hermes} Collaboration}
\affiliation{DESY -- HERMES, Notkestra\ss e 85, D-22607 Hamburg}
\emailAdd{management@hermes.desy.de}

\abstract{
The beam-helicity asymmetry in associated electroproduction of real photons, 
$ep\rightarrow e\gamma \pi N$, in the $\Delta$(1232)-resonance region
is measured using the longitudinally polarized {\sc Hera} positron beam and
an unpolarized hydrogen target.
Azi\-mu\-thal Fourier amplitudes of this asymmetry are extracted separately for two
channels, $ep\rightarrow e\gamma \pi^0 p$ and $ep\rightarrow e\gamma \pi^+ n $,
from a data set collected with a recoil detector.
All asymmetry amplitudes are found to be consistent with zero.
}


\notoc

\begin{document}

\maketitle\flushbottom\newpage

\pagestyle{myplain}
\pagenumbering{roman}

%\section*{The Hermes collaboration}
%\addcontentsline{toc}{section}{The Hermes collaboration}

\section*{The {\usefont{T1}{cmr}{bx}{sc}\selectfont Hermes} Collaboration}

% authors
{%
\begin{flushleft} 
\bf
\hermesauthor[13,16]{A.~Airapetian},
\hermesauthor[27]{N.~Akopov},
\hermesauthor[7]{E.C.~Aschenauer},
\hermesauthor[26]{W.~Augustyniak},
\hermesauthor[27]{R.~Avakian},
\hermesauthor[27]{A.~Avetissian},
\hermesauthor[6]{E.~Avetisyan},
\hermesauthor[18,25]{H.P.~Blok},
\hermesauthor[7]{H.~B\"ottcher},
\hermesauthor[6]{A.~Borissov},
\hermesauthor[14]{J.~Bowles},
\hermesauthor[13]{I.~Brodski},
\hermesauthor[20]{V.~Bryzgalov},
\hermesauthor[14]{J.~Burns},
\hermesauthor[11]{G.P.~Capitani},
\hermesauthor[22]{E.~Cisbani},
\hermesauthor[10]{G.~Ciullo},
\hermesauthor[10]{M.~Contalbrigo},
\hermesauthor[10]{P.F.~Dalpiaz},
\hermesauthor[6]{W.~Deconinck},
\hermesauthor[2]{R.~De~Leo},
\hermesauthor[11]{E.~De~Sanctis},
\hermesauthor[15,9]{M.~Diefenthaler},
\hermesauthor[11]{P.~Di~Nezza},
\hermesauthor[13]{M.~D\"uren},
\hermesauthor[13]{M.~Ehrenfried},
\hermesauthor[27]{G.~Elbakian},
\hermesauthor[5]{F.~Ellinghaus},
\hermesauthor[13]{E.~Etzelm\"uller},
\hermesauthor[7]{R.~Fabbri},
\hermesauthor[22]{S.~Frullani},
\hermesauthor[20]{G.~Gapienko},
\hermesauthor[20]{V.~Gapienko},
\hermesauthor[4]{J.~Garay~Garc\'ia},
\hermesauthor[22]{F.~Garibaldi},
\hermesauthor[6,19,23]{G.~Gavrilov},
\hermesauthor[27]{V.~Gharibyan},
\hermesauthor[15,10]{F.~Giordano},
\hermesauthor[16]{S.~Gliske},
\hermesauthor[6]{M.~Hartig},
\hermesauthor[11]{D.~Hasch},
\hermesauthor[6]{Y.~Holler},
\hermesauthor[7]{I.~Hristova},
\hermesauthor[20]{A.~Ivanilov},
\hermesauthor[1]{H.E.~Jackson},
\hermesauthor[12,15]{S.~Joosten},
\hermesauthor[14]{R.~Kaiser},
\hermesauthor[27]{G.~Karyan},
\hermesauthor[14,13]{T.~Keri},
\hermesauthor[5]{E.~Kinney},
\hermesauthor[19]{A.~Kisselev},
\hermesauthor[20]{V.~Korotkov},
\hermesauthor[17]{V.~Kozlov},
\hermesauthor[9,19]{P.~Kravchenko},
\hermesauthor[8]{V.G.~Krivokhijine},
\hermesauthor[2]{L.~Lagamba},
\hermesauthor[18]{L.~Lapik\'as},
\hermesauthor[14]{I.~Lehmann},
\hermesauthor[10]{P.~Lenisa},
\hermesauthor[16]{W.~Lorenzon},
\hermesauthor[6]{X.-G.~Lu},
\hermesauthor[3]{B.-Q.~Ma},
\hermesauthor[14]{D.~Mahon},
\hermesauthor[15]{N.C.R.~Makins},
\hermesauthor[19]{S.I.~Manaenkov},
\hermesauthor[3]{Y.~Mao},
\hermesauthor[26]{B.~Marianski},
\hermesauthor[27]{H.~Marukyan},
\hermesauthor[23]{C.A.~Miller},
\hermesauthor[24]{Y.~Miyachi},
\hermesauthor[10]{A.~Movsisyan},
\hermesauthor[11]{V.~Muccifora},
\hermesauthor[14]{M.~Murray},
\hermesauthor[6,9]{A.~Mussgiller},
\hermesauthor[19]{Y.~Naryshkin},
\hermesauthor[9]{A.~Nass},
\hermesauthor[7]{M.~Negodaev},
\hermesauthor[7]{W.-D.~Nowak},
\hermesauthor[10]{L.L.~Pappalardo},
\hermesauthor[13]{R.~Perez-Benito},
\hermesauthor[27]{A.~Petrosyan},
\hermesauthor[1]{P.E.~Reimer},
\hermesauthor[11]{A.R.~Reolon},
\hermesauthor[15,7]{C.~Riedl},
\hermesauthor[9]{K.~Rith},
\hermesauthor[14]{G.~Rosner},
\hermesauthor[6]{A.~Rostomyan},
\hermesauthor[1,15]{J.~Rubin},
\hermesauthor[12]{D.~Ryckbosch},
\hermesauthor[20]{Y.~Salomatin},
\hermesauthor[21]{A.~Sch\"afer},
\hermesauthor[4,12]{G.~Schnell},
\hermesauthor[14]{B.~Seitz},
\hermesauthor[24]{T.-A.~Shibata},
\hermesauthor[13]{M.~Stahl},
\hermesauthor[10]{M.~Statera},
\hermesauthor[9]{E.~Steffens},
\hermesauthor[18]{J.J.M.~Steijger},
\hermesauthor[7]{J.~Stewart},
\hermesauthor[9]{F.~Stinzing},
\hermesauthor[27]{S.~Taroian},
\hermesauthor[17]{A.~Terkulov},
\hermesauthor[15]{R.~Truty},
\hermesauthor[26]{A.~Trzcinski},
\hermesauthor[12]{M.~Tytgat},
\hermesauthor[12]{Y.~Van~Haarlem},
\hermesauthor[4,12]{C.~Van~Hulse},
\hermesauthor[19]{V.~Vikhrov},
\hermesauthor[2]{I.~Vilardi},
\hermesauthor[3]{S.~Wang},
\hermesauthor[6,7,9]{S.~Yaschenko},
\hermesauthor[6]{Z.~Ye},
\hermesauthor[23]{S.~Yen},
\hermesauthor[6,13]{V.~Zagrebelnyy},
\hermesauthor[6]{B.~Zihlmann},
\hermesauthor[26]{P.~Zupranski}
\end{flushleft} 
}
%-- HERMES Institutes
\bigskip
{\it
\begin{flushleft} 
\hermesinstitute[1]{Physics Division, Argonne National Laboratory, Argonne, Illinois 60439-4843, USA}
\hermesinstitute[2]{Istituto Nazionale di Fisica Nucleare, Sezione di Bari, 70124 Bari, Italy}
\hermesinstitute[3]{School of Physics, Peking University, Beijing 100871, China}
\hermesinstitute[4]{Department of Theoretical Physics, University of the Basque Country UPV/EHU, 48080 Bilbao, Spain and IKERBASQUE, Basque Foundation for Science, 48011 Bilbao, Spain}
\hermesinstitute[5]{Nuclear Physics Laboratory, University of Colorado, Boulder, Colorado 80309-0390, USA}
\hermesinstitute[6]{DESY, 22603 Hamburg, Germany}
\hermesinstitute[7]{DESY, 15738 Zeuthen, Germany}
\hermesinstitute[8]{Joint Institute for Nuclear Research, 141980 Dubna, Russia}
\hermesinstitute[9]{Physikalisches Institut, Universit\"at Erlangen-N\"urnberg, 91058 Erlangen, Germany}
\hermesinstitute[10]{Istituto Nazionale di Fisica Nucleare, Sezione di Ferrara and Dipartimento di Fisica e Scienze della Terra, Universit\`a di Ferrara, 44122 Ferrara, Italy}
\hermesinstitute[11]{Istituto Nazionale di Fisica Nucleare, Laboratori Nazionali di Frascati, 00044 Frascati, Italy}
\hermesinstitute[12]{Department of Physics and Astronomy, Ghent University, 9000 Gent, Belgium}
\hermesinstitute[13]{II. Physikalisches Institut, Justus-Liebig Universit\"at Gie{\ss}en, 35392 Gie{\ss}en, Germany}
\hermesinstitute[14]{SUPA, School of Physics and Astronomy, University of Glasgow, Glasgow G12 8QQ, United Kingdom}
\hermesinstitute[15]{Department of Physics, University of Illinois, Urbana, Illinois 61801-3080, USA}
\hermesinstitute[16]{Randall Laboratory of Physics, University of Michigan, Ann Arbor, Michigan 48109-1040, USA }
\hermesinstitute[17]{Lebedev Physical Institute, 117924 Moscow, Russia}
\hermesinstitute[18]{National Institute for Subatomic Physics (Nikhef), 1009 DB Amsterdam, The Netherlands}
\hermesinstitute[19]{B.P. Konstantinov Petersburg Nuclear Physics Institute, Gatchina, 188300 Leningrad Region, Russia}
\hermesinstitute[20]{Institute for High Energy Physics, Protvino, 142281 Moscow Region, Russia}
\hermesinstitute[21]{Institut f\"ur Theoretische Physik, Universit\"at Regensburg, 93040 Regensburg, Germany}
\hermesinstitute[22]{Istituto Nazionale di Fisica Nucleare, Sezione di Roma, Gruppo Collegato Sanit\`a and Istituto Superiore di Sanit\`a, 00161 Roma, Italy}
\hermesinstitute[23]{TRIUMF, Vancouver, British Columbia V6T 2A3, Canada}
\hermesinstitute[24]{Department of Physics, Tokyo Institute of Technology, Tokyo 152, Japan}
\hermesinstitute[25]{Department of Physics and Astronomy, VU University, 1081 HV Amsterdam, The Netherlands}
\hermesinstitute[26]{National Centre for Nuclear Research, 00-689 Warsaw, Poland}
\hermesinstitute[27]{Yerevan Physics Institute, 375036 Yerevan, Armenia}
\end{flushleft} 
}

\clearpage
\pagenumbering{arabic}

\noindent\rule\textwidth{.1pt}
\tableofcontents
\afterTocSpace
\noindent\rule\textwidth{.1pt}

\pagestyle{myplain}

%%%%%%%%%%%%%%%%%%%%%%%%%%%%%%%%%%%%%%%%%%%%%%%%%%%%%%%%%%%%%%%%%%%%%%%%%%%%%%%%%%%%%%%%%%%%%%%%%%%
%%%%%%%%%%%%%%%%%%%%%%%%%%%%%%%%%%%%%%%%%%%%%%%%%%%%%%%%%%%%%%%%%%%%%%%%%%%%%%%%%%%%%%%%%%%%%%%%%%%
%%%%%%%%%%%%%%%%%%%%%%%%%%%%%%%%%%%%%%%%%%%%%%%%%%%%%%%%%%%%%%%%%%%%%%%%%%%%%%%%%%%%%%%%%%%%%%%%%%%

\section{Introduction}
\label{sec:Introduction}

There continues to be intense interest in Generalized Parton Distributions
(GPDs) \cite{Mueller, Radyushkin, Ji}, both theoretical and experimental.
These distributions relate to the total angular
momentum of partons in the nucleon \cite{Ji_2} and information
on the parton's transverse location in the nucleon correlated
with the fraction of the nucleon's longitudinal momentum carried by that parton \cite{Burkardt}.
The GPDs that have thus far attracted the most interest parametrize the nonperturbative part
of hard exclusive reactions where  the target system stays intact
such as $ep \rightarrow e\gamma p$. They
depend on four kinematic variables: $t$, $x$, $\xi$, and $Q^2$. 
The Mandelstam variable $t=(p-p^{\prime})^2$ is the square of the difference between
the initial ($p$) and final ($p^{\prime}$) four-momenta of the target nucleon.
The variable $x$ is the average of the initial and final fractions of the (large)
target longitudinal momentum that is carried by the struck parton, and the variable $\xi$,
known as the skewness, is half of the difference between these fractions. 
The evolution of GPDs with the photon virtuality $Q^2\equiv -q^2$ is analogous
to that of parton distribution functions, with $q=k-k^{\prime}$ being the difference
between the  four-momenta of the incident and the scattered leptons. 
Currently, no hard exclusive measurements exist that provide access to $x$.
Because of the lack of consensus about the definition of $\xi$ in terms of experimental
observables, the results are typically reported by {\sc Hermes} as projections
in $x_{\textrm{B}}\equiv Q^2/(2pq)$, to which $\xi$ can be related through
$\xi\simeq x_{\textrm{B}}/(2-x_{\textrm{B}})$ in the generalized Bjorken limit of
large $Q^2$ and fixed $x_{\textrm{B}}$ and $t$.
Several GPDs describe various possible helicity transitions of the struck quark
and/or the nucleon.
At leading twist (i.e., twist-2) and for a spin-1/2 target such as the proton,
four chiral-even GPDs ($H^{q}$, $\widetilde{H}^{q}$, $E^{q}$, $\widetilde{E}^{q}$) are required
to describe processes that conserve the helicity of the struck quark with flavor $q$. 

The GPD formalism can be extended to more general baryonic final states, 
in particular here to the $\Delta$ resonance.
Similar to  $N \rightarrow \Delta$ ``transition" form factors, one can
introduce $N \rightarrow \Delta$ transition GPDs.
At leading twist, the $\gamma^*N \rightarrow \gamma\Delta$ process can be parametrized
in terms of three vector and four axial-vector $N \rightarrow \Delta$ GPDs~\cite{Fra00}.
Among them, one expects three such GPDs to dominate at small $|t|$:
the magnetic vector GPD $H_M$, of which the first moment corresponds
to the $N \rightarrow \Delta$ magnetic dipole transition form factor $G_M^*(t)$, and the
axial-vector GPDs $C_1$ and $C_2$, of which the first moments
correspond to the axial-vector and pseudoscalar $N \rightarrow \Delta$ form factors,
respectively.

In ref.~\cite{Asso03}, a model is proposed to describe the ``associated'' reaction
$ep\rightarrow e\gamma \pi N$.
In this model, the so-called soft-pion technique that is based on current algebra
and chiral symmetry allows for S-wave pions
the use of the same GPDs as in $ep \rightarrow e\gamma p$.
In order to extend the model estimations to pions of higher energy, 
the P-wave production is assumed to be dominated by the $\Delta (1232)$
isobar production and is added following the large $N_c$ limit approach for $N \rightarrow \Delta$
GPDs developed in refs.~\cite{Fra00, GPV01}.
In this model, the $N \rightarrow \Delta$ GPDs $H_{M}$, $C_{1}$,
and $C_{2}$ are connected to the $N\rightarrow N$ isovector GPDs as:

\begin{eqnarray}
H_{M}(x,\xi ,t) & = & 
\frac{2}{\sqrt{3}}\left[ E^{u}(x,\xi , t)
-E^{d}(x,\xi ,t)\right],
\nonumber \\
C_{1}(x,\xi , t) & = & 
\sqrt{3}\left[ \widetilde{H}^{u}(x,\xi , t)
-\widetilde{H}^{d}(x,\xi , t)\right],\nonumber \\
C_{2}(x,\xi , t) & = & 
\frac{\sqrt{3}}{4}\left[ \widetilde{E}^{u}(x,\xi , t)
-\widetilde{E}^{d}(x,\xi , t)\right].
\label{Eq_DEL.8} 
\end{eqnarray}

\noindent This estimate is expected to have an accuracy of about 30\%.
Thus, these large $N_c$ relations allow the interpretation of 
the associated reaction in terms of nucleon GPDs and therefore open (model-dependent)
access to different flavor combinations of the nucleon GPDs.
For example, $ep \rightarrow e\gamma p$ is sensitive to the combination
$\frac{4}{9}\widetilde H^u+\frac{1}{9}\widetilde H^d$, whereas in $ep \rightarrow e\gamma \Delta$
the isovector part $\widetilde H^u -\widetilde H^d$ appears.

As for the $ep \rightarrow e\gamma p$ reaction, for the associated reaction
the amplitudes of the Deeply Virtual Compton Scattering (DVCS) process and
of the Bethe--Heitler (BH) process,
in which a bremsstrahlung photon is radiated from the incident or scattered lepton,
combine coherently.
In the absence of available data for the associated reaction,
the pion photoproduction cross section calculated using
an approach similar to that applied to the associated BH process is
compared in ref.~\cite{Asso03} with
experimental data from refs.~\cite{Ref36, Ref37, Ref38}. Around the $\Delta$-resonance mass,
the model overestimates the experimental cross sections by about 10\%.

In this paper, the first measurement of the single-charge beam-helicity asymmetry
in the reaction $ep \rightarrow e\gamma \pi N$ is presented and compared with
model predictions. The asymmetry is defined as in ref.~\cite{PublicationDraft92} to be

\begin{eqnarray}
\mathcal{A}_{\mathrm{LU}}(\phi, e_\ell)
&=&
\frac{\sigma_{\mathrm{LU}}(\phi, e_\ell,\lambda=+1)-\sigma_{\mathrm{LU}}(\phi, e_\ell,\lambda=-1)}{\sigma_{\mathrm{LU}}(\phi,e_\ell,\lambda=+1)+\sigma_{\mathrm{LU}}(\phi,e_\ell,\lambda=-1)}.
\label{eq:alu_sbc2}
\end{eqnarray}

\noindent Here, $\sigma_{\mathrm{LU}}$ denotes the differential cross section
for longitudinally polarized beam and unpolarized target,
$\lambda=\pm 1$ and $e_\ell(=+ 1)$ are respectively the helicity and unit charge
of the beam lepton, and the angle $\phi$ is the azimuthal orientation
of the photon production plane with respect to the lepton scattering plane.
The definition of the angle $\phi$ follows the Trento conventions \cite{Tre04}.
The asymmetries are extracted in the kinematic range of $-t < 1.2$\,GeV$^2$,
$0.03 < x_{B} < 0.35$, and $1$\,GeV$^2<Q^2<10$\,GeV$^2$.

%%%%%%%%%%%%%%%%%%%%%%%%%%%%%%%%%%%%%%%%%%%%%%%%%%%%%%%%%%%%%%%%%%%%%%%%%%%%%%%%%%%%%%%%%%%%%%%%%%%
%%%%%%%%%%%%%%%%%%%%%%%%%%%%%%%%%%%%%%%%%%%%%%%%%%%%%%%%%%%%%%%%%%%%%%%%%%%%%%%%%%%%%%%%%%%%%%%%%%%
%%%%%%%%%%%%%%%%%%%%%%%%%%%%%%%%%%%%%%%%%%%%%%%%%%%%%%%%%%%%%%%%%%%%%%%%%%%%%%%%%%%%%%%%%%%%%%%%%%%

\section{The {\usefont{T1}{cmr}{bx}{sc}\selectfont Hermes} experiment in 2006--2007}
\label{sec:experiment}

The data presented here were collected in 2006 and 2007 at {\sc Hermes} ({\sc Desy})
using the 27.6 GeV {\sc Hera} positron beam and an unpolarized hydrogen gas target internal
to the beam line. For this measurement, the recoil detector \cite{RecoilTechnicalPaper}
was used in conjunction with the forward spectrometer \cite{hermes:spectrometer}.

The {\sc Hera} lepton beam was transversely self-polarized by the emission of synchrotron
radiation \cite{Sokolov+:1964}.
Longitudinal polarization of the beam in the target region was achieved by a pair
of spin rotators located upstream and downstream of the experiment \cite{Buon:1986}.
The sign of the beam polarization was reversed three times over the running period.
Two Compton backscattering polarimeters \cite{TPOL:1994, LPOL:2002} independently
measured the longitudinal and transverse beam polarizations.

For the analysis of the beam-helicity asymmetry considered here,
data collected with only one lepton beam charge
($e_\ell=+1$) and both beam-helicity states are available.
For this data set, the average beam polarization was $P_\ell=0.402$ ($-0.394$)
for positive (negative) beam helicity, with a total relative uncertainty
of 1.96\% \cite{HERApol2012}.

The scattered lepton and particles produced in the polar-angle range
$0.04\;\mathrm{rad}<\theta<0.22\;\mathrm{rad}$ were detected by the
forward spectrometer, for which the average lepton-identification efficiency
was at least 98\% with hadron contamination of less than 1\%.
The produced particles emerging at large polar angles and with small momenta were
detected by the recoil detector in the polar-angle range
$0.25\;\mathrm{rad}<\theta<1.45\;\mathrm{rad}$, with an azimuthal coverage of about 75\%.
The lower-momentum detection threshold for protons (pions) was 125 (60) MeV for this analysis.

The recoil detector surrounded the target cell and consisted of
a Silicon Strip Detector (SSD), a Scintillating Fiber Tracker (SFT),
and a photon detector, all
embedded in a solenoidal magnetic field with field strength of 1\,T.
A detailed description of the recoil-detector components is given in
ref.~\cite{RecoilTechnicalPaper}.

Track search and momentum reconstruction in the recoil detector
are performed by combining coordinate information from the SSD and SFT layers.
For protons, energy deposition in the SSD is additionally taken into account.
This improves the momentum resolution for momenta below 0.5\,GeV, leading to
a resolution of 2-10\% from 0.15\,GeV to 0.5\,GeV
\cite{RecoilTechnicalPaper}.
For pions, the momentum resolution is about 12\% and almost independent of momentum.
The azimuthal- and polar-angle resolution is about 4\,mrad and 10\,mrad respectively
for pions and for protons with momenta larger than 0.5\,GeV,
deteriorating for lower proton momenta because of multiple scattering.

For each reconstructed track, the energy deposited along the particles'
trajectory through the active detector components is used to determine the particle type.
As protons and pions dominate the event sample, only the separation
of these two particle types is considered.
For each detection layer $i$, a particle-identification discriminator
$\mathrm{rdPID}_{i}$, which depends on the reconstructed three-momentum $|\vec p|$ and 
on the energy deposition $dE$ normalized to pathlength,
is calculated according to

\begin{equation}
\mathrm{rdPID}_{i}(dE; |\vec p|) =
\log_{10}\frac{D_{i}(dE; \beta\gamma = \frac{|\vec p|}{M_{p}}) }
{D_{i}(dE; \beta\gamma = \frac{|\vec p|}{M_{\pi}}) },
\label{eq:PID_prob}
\end{equation}

\noindent
where the ``parent distributions'' $D_{i}$ are energy-deposition distributions
normalized to unity, $\beta$ is the ratio of the particle velocity to the speed of light,
$\gamma$ is the Lorentz factor, and $M_{p}$ ($M_{\pi}$) is the proton (pion) mass.
The combined particle-identification discriminator rdPID is the sum of the
discriminators $\mathrm{rdPID}_{i}$ from the individual layers.
A constraint on rdPID is chosen to distinguish between charged pions and
protons, while providing an appropriate compromise between efficiency
and contamination \cite{Xianguo}.

Details of the tracking, momentum reconstruction, and particle-identification procedures
as well as detector performance studies are presented in ref.~\cite{RecoilTechnicalPaper}.

%%%%%%%%%%%%%%%%%%%%%%%%%%%%%%%%%%%%%%%%%%%%%%%%%%%%%%%%%%%%%%%%%%%%%%%%%%%%%%%%%%%%%%%%%%%%%%%%%%%
%%%%%%%%%%%%%%%%%%%%%%%%%%%%%%%%%%%%%%%%%%%%%%%%%%%%%%%%%%%%%%%%%%%%%%%%%%%%%%%%%%%%%%%%%%%%%%%%%%%
%%%%%%%%%%%%%%%%%%%%%%%%%%%%%%%%%%%%%%%%%%%%%%%%%%%%%%%%%%%%%%%%%%%%%%%%%%%%%%%%%%%%%%%%%%%%%%%%%%%

\section{Event selection}
\label{sec:eventselection}

A positron trigger is formed from a coincidence between three scintillator
hodoscope planes and a lead-glass calorimeter.
Following the approach of ref. \cite{BCA2007}, inclusive $ep\rightarrow eX$
events in the Deep-Inelastic Scattering (DIS) regime are selected by imposing the
following kinematic requirements on the identified positron with the largest momentum
in the event:
$Q^2>1$\,GeV$^2$, $W^2>9$\,GeV$^2$, and $\nu<22$\,GeV,
where $W$ is the invariant mass of the $\gamma^*p$ system and $\nu\equiv(pq)/M_p$ 
the energy of the virtual photon in the target-rest frame.
This sample of inclusive DIS events is employed for the determination of relative
luminosities of the two beam-helicity states as inclusive DIS with virtual-photon exchange
from unpolarized targets is invariant under reversal of the beam helicity.

Exclusive $ep\rightarrow e\gamma \pi N$ event candidates are selected from the DIS sample
by requiring in the forward spectrometer the detection of exactly one identified positron
in the absence of other charged particles and of exactly one signal cluster in the calorimeter
not associated with the positron and hence signifying a real photon.
The kinematic requirements on the identified positron and the photon cluster
applied in ref.~\cite{PublicationDraft92} are adjusted for this analysis
as follows in order to optimize the selection of $ep \rightarrow e\gamma \pi N$ events.
The cluster is required to represent an energy deposition above 8\,GeV in the calorimeter
and above 1\,MeV in the preshower detector.
Two kinematic constraints are applied: the polar angle $\theta_{\gamma^*\gamma}$ between the laboratory
three-momenta of the virtual and real photons is limited to be less than 70\,mrad,
and the value of $-t$ is limited to be less than 1.2\,GeV$^2$.
Here, $-t$ is calculated without the use of either the photon-energy
measurement or recoil-detector
information, under the hypothesis of an exclusive $ep\rightarrow e\gamma \Delta^+$ event.
(The width of the $\Delta^+$ is small compared to the experimental resolution.)

All recoil tracks identified as protons and positively charged pions are considered
in order to select the associated reactions $ep\rightarrow e\gamma \pi^0 p$ and
$ep\rightarrow e\gamma \pi^+ n$ in the $\Delta$-resonance region. Kinematic
event fitting is performed under the corresponding hypotheses using the three-momenta
of the positron and photon measured in the forward spectrometer and the proton
(pion) track in the recoil detector.
The neutral pion (neutron) is not identified, therefore
the fit enforces two four-momentum conservation equations based on 
the assumption of the $ep\rightarrow e\gamma \Delta^+$ reaction
with $\Delta^+$ decay to $p\pi^0 (n\pi^+)$ assuming the PDG value of the $\Delta^+$(1232) mass.
In addition, adopting $\pi^+$ as proton candidates,
the kinematic fit described in ref.~\cite{PublicationDraft92} is performed in order
to suppress $ep\rightarrow e\gamma p$ background events.
The following constraints on the $\chi^2$ of kinematic event fitting and
on the rdPID values are optimized and applied for the selection of
events from the associated channels:
\begin{itemize}
\item $ep\rightarrow e\gamma \pi^0 p$: \ $\chi_{ep\rightarrow e\gamma \pi^0 p}^2 < 4.6$, $\chi_{ep\rightarrow e\gamma p}^2 > 50$, and rdPID~$>0$ (to select protons),
\item $ep\rightarrow e\gamma \pi^+ n$: \ $\chi_{ep\rightarrow e\gamma \pi^+ n}^2 < 4.6$, $\chi_{ep\rightarrow e\gamma p}^2 > 50$, and rdPID~$<0$ (to select pions).
\end{itemize}

Kinematic distributions obtained from experimental data are compared with a mixture of simulated
data samples.
Following the approach of
refs. \cite{PublicationDraft69, PublicationDraft90, PublicationDraft92},
BH events are simulated using the Mo--Tsai formalism \cite{motsai}, by an event generator
based on ref.~\cite{gmcDVCS} and described in detail in ref.~\cite{BernhardsPhD}.
This sample of BH events includes events from associated production generated using
the parametrization of the form factor for the resonance region from ref.~\cite{Bra76}.
The individual cross sections for single-meson decay channels of $\Delta^+$
are treated according to the MAID2000 model~\cite{MAID}.
(Neither the DVCS process nor the associated DVCS process are included in the
simulation since for the latter an event generator is unavailable.)
Semi-Inclusive DIS (SIDIS) events are simulated using an event generator based
on LEPTO \cite{lepto} with a set of JETSET~\cite{jetset} fragmentation parameters
tuned for HERMES kinematic conditions \cite{Achim},
including the RADGEN \cite{radgen} package for radiative effects.

\begin{figure}[h!]
\centerline{
\hspace{0.2cm}
\epsfig{file=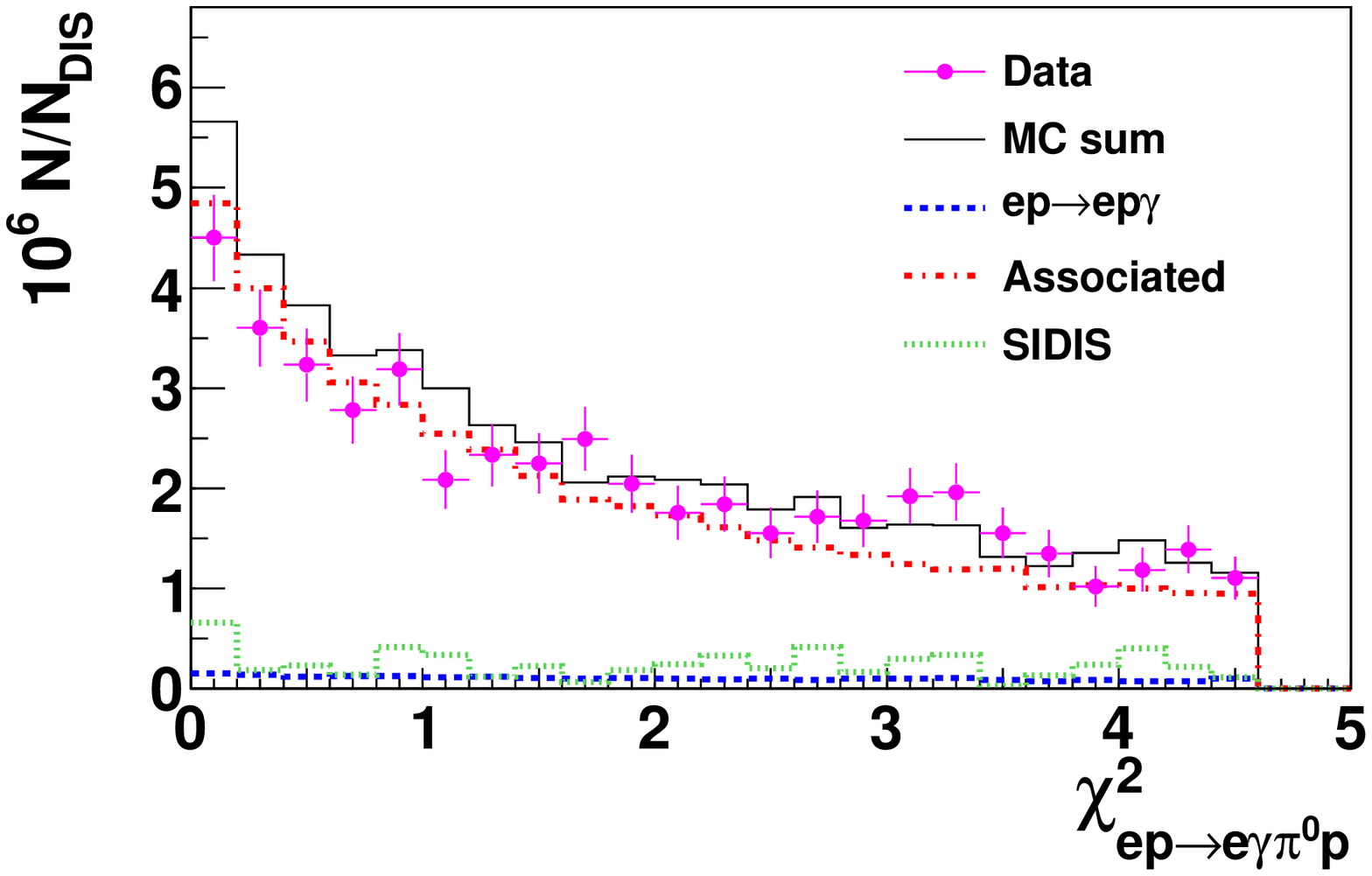, width=0.525\textwidth}
\hspace{-0.55cm}
\epsfig{file=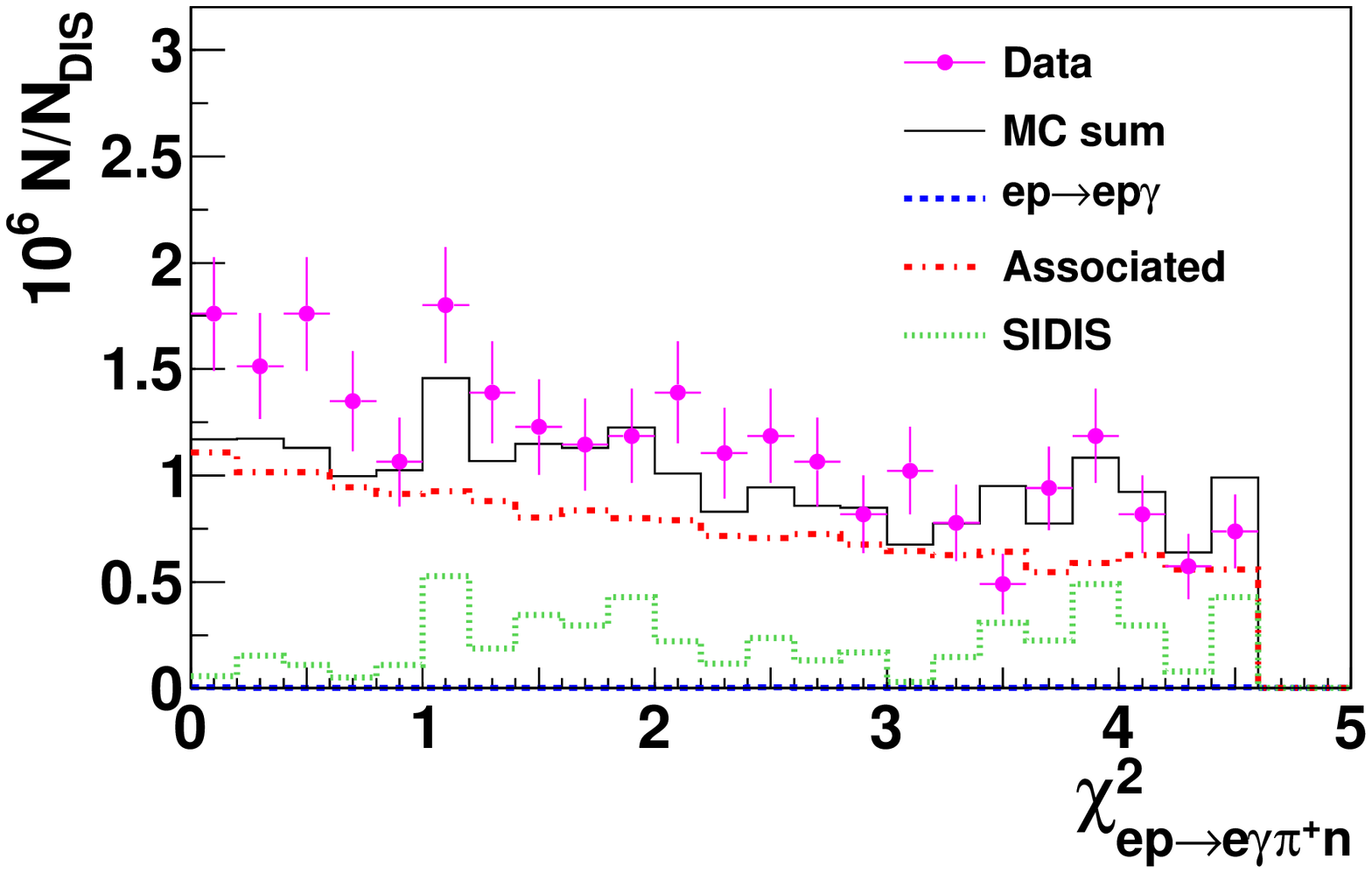, width=0.525\textwidth}
}
\caption{\label{fig:asso_chi_6_8} Distributions of $\chi_{ep\rightarrow e\gamma \pi^0 p}^2$ (left)
and $\chi_{ep\rightarrow e\gamma \pi^+ n}^2$ (right) for the channel $ep\rightarrow e\gamma \pi^0 p$
and $ep\rightarrow e\gamma \pi^+ n$, respectively.
Experimental data are presented by points, and the results of the
Monte Carlo simulation by lines.
Contributions from the associated, $ep\rightarrow e\gamma p$,
and SIDIS reactions are shown by red dash-dotted, blue dashed, and green dotted lines,
respectively (color online).
Data and Monte Carlo yields are normalized to the corresponding numbers of DIS events. 
}
\end{figure}

The $\chi^2$ distributions from kinematic fitting
under the hypothesis of the associated reaction obtained for experimental
and simulated data are compared in figure~\ref{fig:asso_chi_6_8}
for the channels $ep\rightarrow e\gamma \pi^0 p$ (left panel)
and $ep\rightarrow e\gamma \pi^+ n$ (right panel).
For both channels, acceptable agreement in the shape of the distributions is observed,
given that the Monte Carlo event generator does not include the DVCS processes.

\begin{figure}[h!]
\centerline{
\hspace{0.2cm}
\epsfig{file=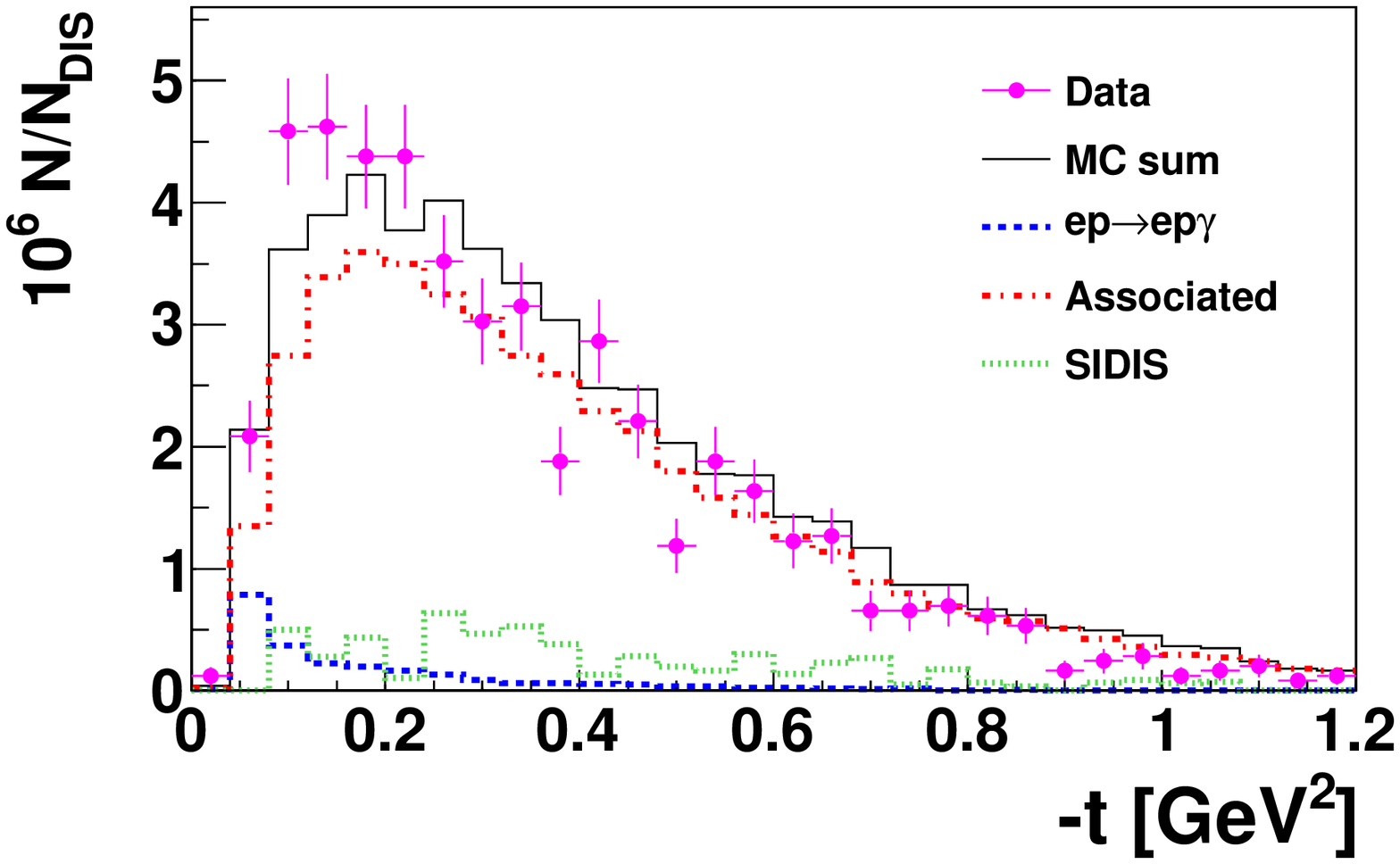, width=0.525\textwidth}
\hspace{-0.55cm}
\epsfig{file=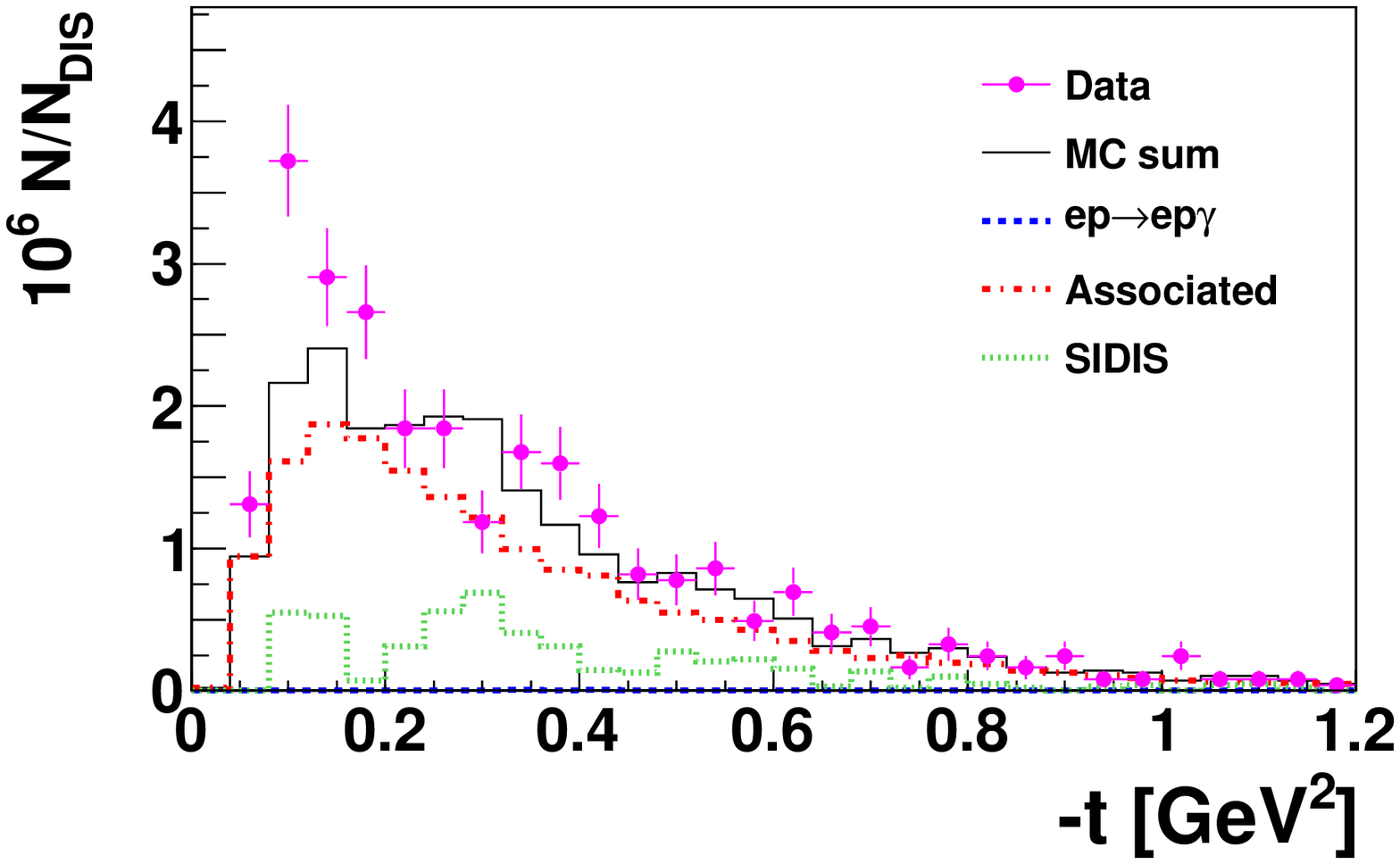, width=0.525\textwidth}
}
\centerline{
\hspace{0.2cm}
\epsfig{file=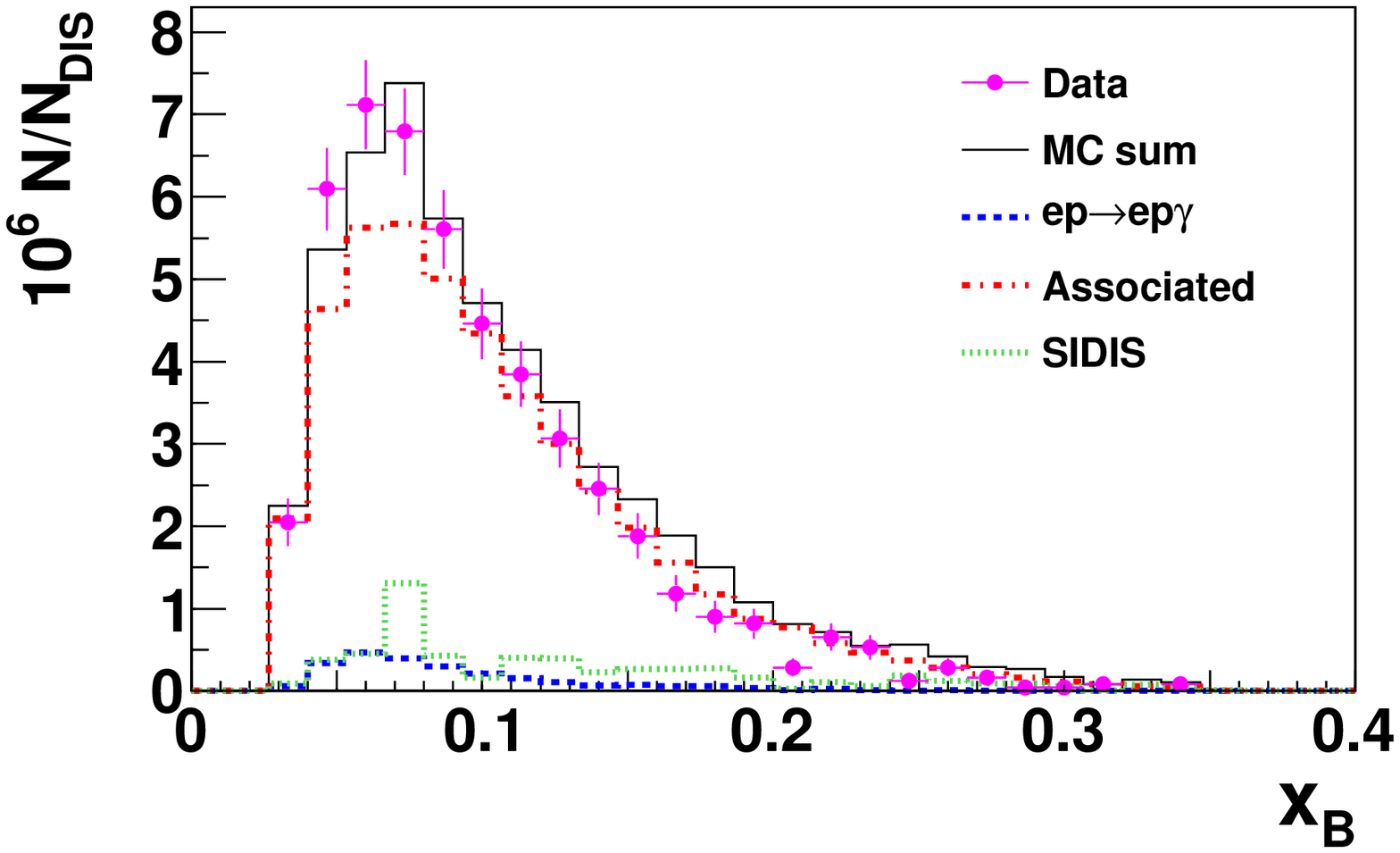, width=0.525\textwidth}
\hspace{-0.55cm}
\epsfig{file=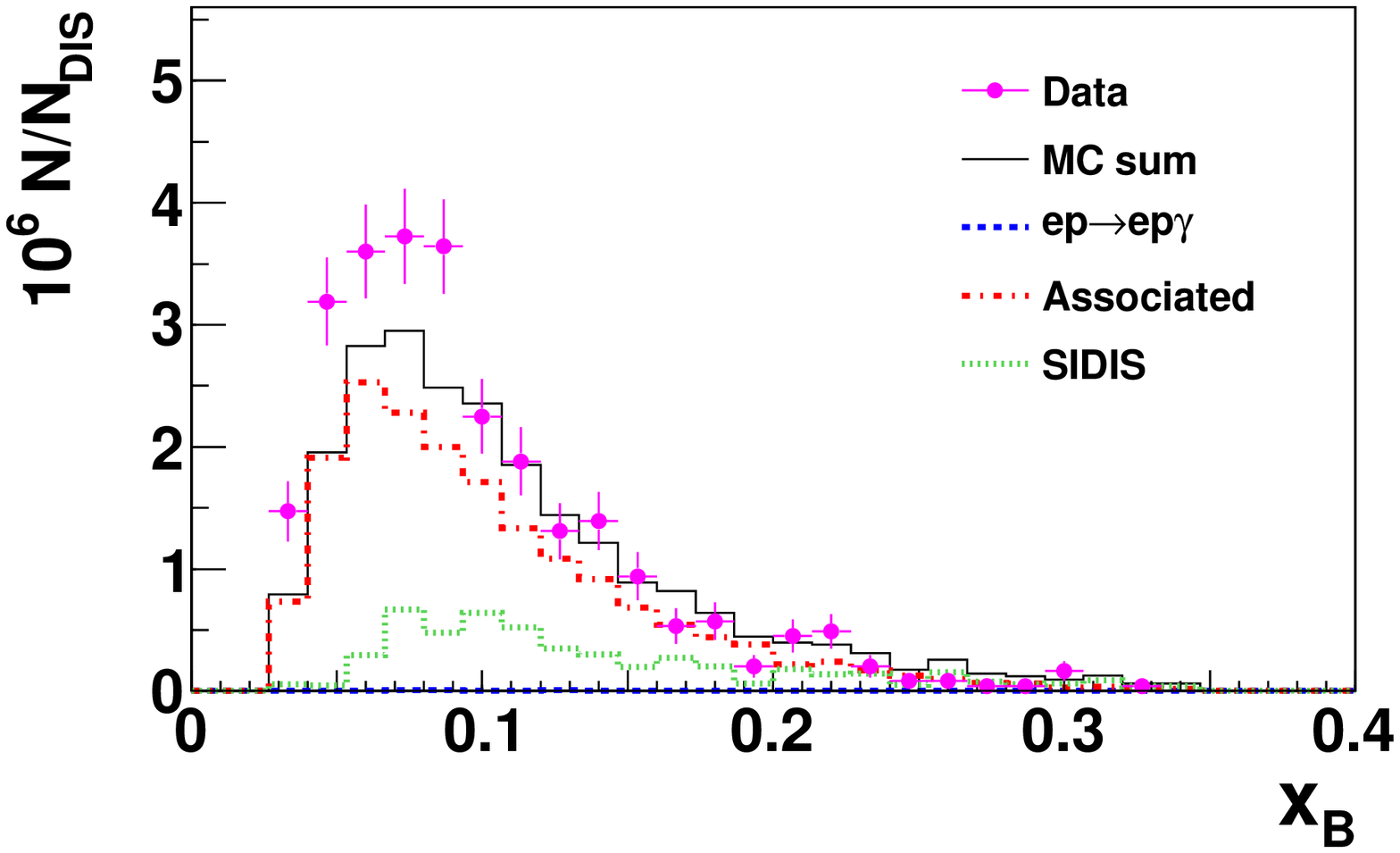, width=0.525\textwidth}
}
\centerline{
\hspace{0.2cm}
\epsfig{file=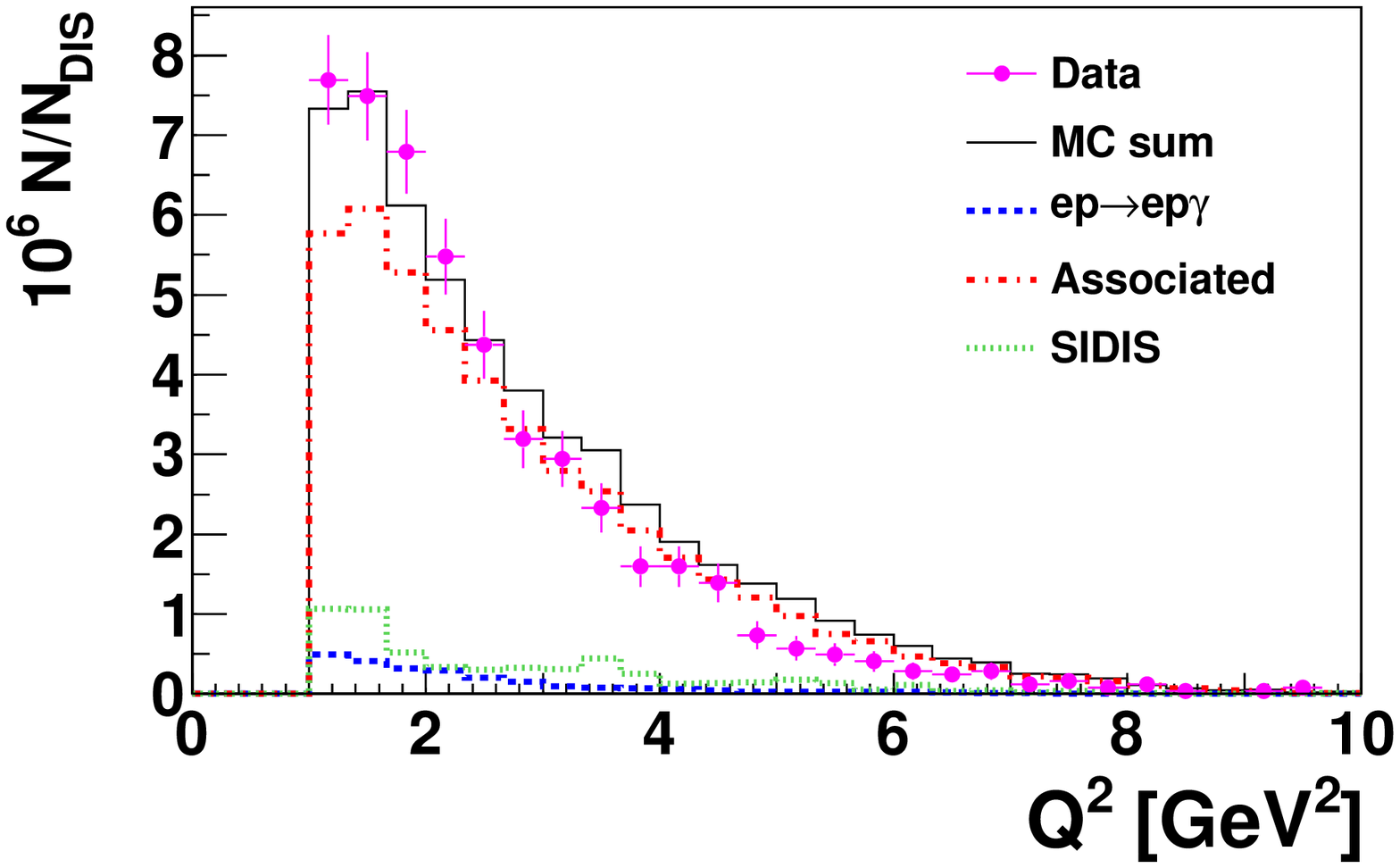, width=0.525\textwidth}
\hspace{-0.55cm}
\epsfig{file=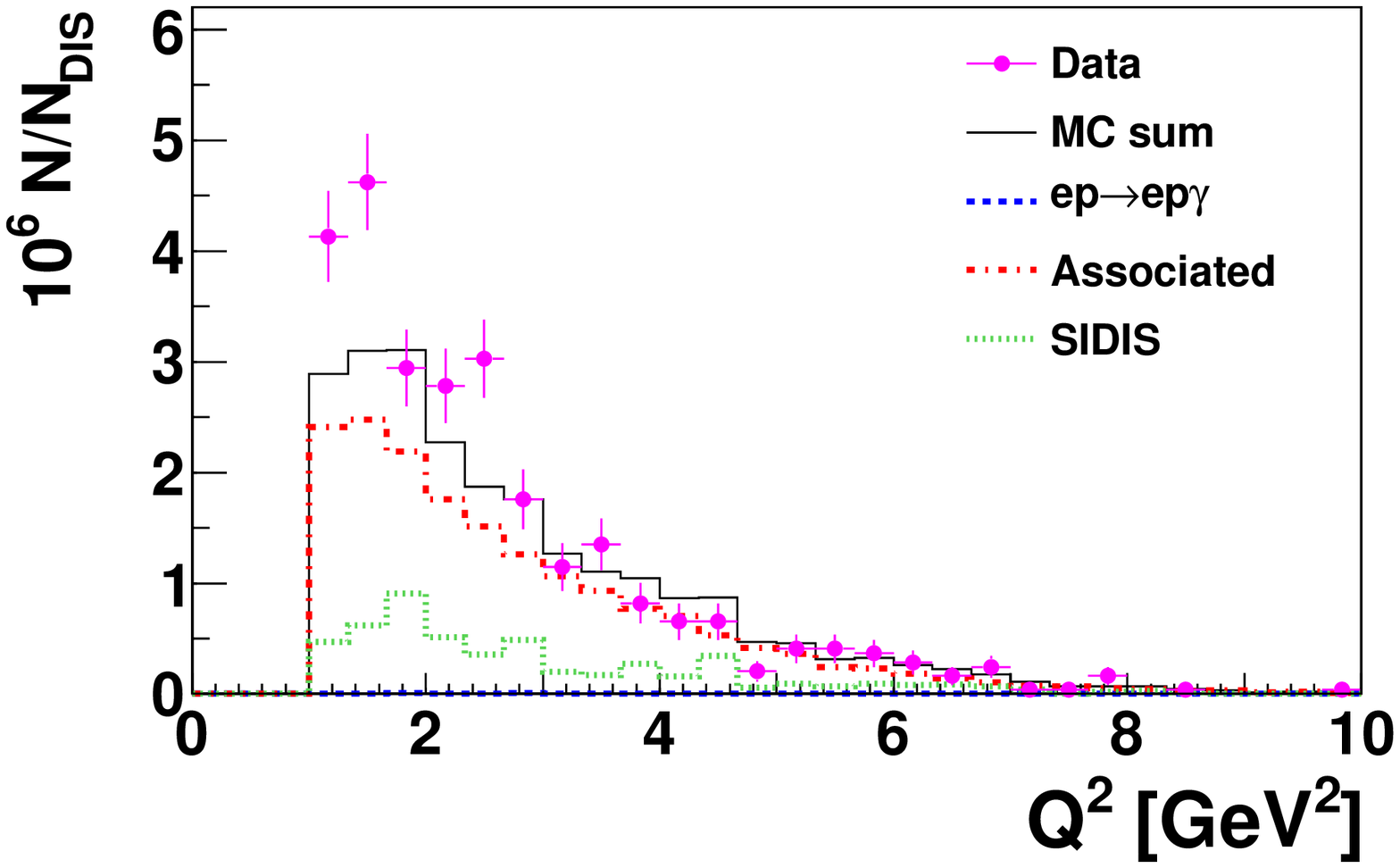, width=0.525\textwidth}
}
\caption{\label{fig:asso_txbq2_6_8} Distributions of $t$ (top row), $x_B$ (middle row),
and $Q^2$ (bottom row) for the associated channel $ep\rightarrow e\gamma \pi^0 p$ (left column) and
$ep\rightarrow e\gamma \pi^+ n$ (right column).
Notations are the same as in figure~\ref{fig:asso_chi_6_8}.
}
\end{figure}

In figure~\ref{fig:asso_txbq2_6_8} comparisons of distributions over the kinematic variables
$-t$, $x_B$, and $Q^2$ are shown for the associated channels $ep\rightarrow e\gamma \pi^0 p$
(left panel) and $ep\rightarrow e \gamma \pi^+ n$ (right panel).
This comparison provides evidence that the Monte Carlo description of the associated BH reaction
used in previous analyses
\cite{BCA2007, PublicationDraft68, PublicationDraft69, PublicationDraft90, PublicationDraft92}
accounts for most of the observed yields.

The fractional contributions from the associated reaction, $ep\rightarrow e\gamma p$,
and SIDIS processes, obtained by analyzing Monte Carlo data in the same
way as described above,
are listed with their statistical uncertainties in table~\ref{tab:frac1}
for the channel $ep\rightarrow e\gamma \pi^0 p$ and in table~\ref{tab:frac2}
for the channel $ep\rightarrow e\gamma \pi^{+} n$ in one kinematic bin covering
the entire kinematic region considered here (``overall'') and in kinematic bins of
$-t$, $x_{\mathrm{B}}$, and $Q^{2}$. 

\begin{table}[h!]
\footnotesize 
\begin{center}
\begin{tabular}{|c|c|c|c|c|}
\hline
 \multicolumn{2}{|c|}{Kinematic bin} & $ep\rightarrow e\gamma \pi^0 p$ [$\%$]& $ep\rightarrow e\gamma p$ [$\%$]& SIDIS [$\%$] \\ \hline\hline
\multicolumn{2}{|c|}{Overall}  & 85 $\pm$ 1 & 4.6 $\pm$ 0.1 & 11 $\pm$ 1  \\\hline\hline
             & $<$0.17   & 79 $\pm$ 2 & 13.5 $\pm$ 0.5 \, & \, 8 $\pm$ 3 \\
 $-t$        & 0.17-0.30 & 86 $\pm$ 3 & 3.9 $\pm$ 0.2 & 11 $\pm$ 3 \\
$[\mathrm{GeV}^{2}]$  & 0.30-0.50 & 86 $\pm$ 2 & 2.1 $\pm$ 0.1 & 12 $\pm$ 2 \\
             & 0.50-1.20 & 86 $\pm$ 2 & 1.3 $\pm$ 0.1 &  13 $\pm$ 2 \\\hline\hline
             & 0.03-0.07 & 86 $\pm$ 2 & 6.3 $\pm$ 0.3 &  \, 8 $\pm$ 2 \\
$x_{B}$     & 0.07-0.10 & 84 $\pm$ 3 & 5.1 $\pm$ 0.2 & 11 $\pm$ 3 \\
             & 0.10-0.15 & 88 $\pm$ 2 & 3.5 $\pm$ 0.2 & \, 9 $\pm$ 2 \\
             & 0.15-0.35 & 79 $\pm$ 2 & 3.1 $\pm$ 0.2 & 18 $\pm$ 2 \\\hline\hline
             & 1.00-1.50 & 78 $\pm$ 3 & 6.3 $\pm$ 0.4 & 16 $\pm$ 4 \\
$Q^{2}$      & 1.50-2.30 & 86 $\pm$ 2 & 5.5 $\pm$ 0.2 & \, 8 $\pm$ 2 \\
$[\mathrm{GeV}^{2}]$  & 2.30-3.50 & 86 $\pm$ 2 & 3.9 $\pm$ 0.2 & 10 $\pm$ 2 \\
             & 3.50-10.0 & 86 $\pm$ 2 & 3.1 $\pm$ 0.2 &  11 $\pm$ 2 \\\hline
 \end{tabular}
\end{center}
\caption{
\label{tab:frac1} Monte-Carlo-estimated fractional contributions
to the measured yields by $\,$
$ep\rightarrow e\gamma \pi^0 p$, $ep\rightarrow e\gamma p$,
and SIDIS reactions in the selected sample of $ep\rightarrow e\gamma \pi^0 p$ events.
}
\end{table}

\begin{table}[h!]
\footnotesize 
\begin{center}
\begin{tabular}{|c|c|c|c|c|}
\hline
 \multicolumn{2}{|c|}{Kinematic bin} & $ep\rightarrow e\gamma \pi^+ n$ [$\%$]& $ep\rightarrow e\gamma p$ [$\%$]& SIDIS [$\%$] \\ \hline\hline
\multicolumn{2}{|c|}{Overall}  & 77 $\pm$ 2 & 0.2 $\pm$ 0.1 & 23 $\pm$ 3  \\\hline\hline
             & $<$0.17   & 82 $\pm$ 5 & 0.1 $\pm$ 0.1 & 18 $\pm$ 5 \\
 $-t$        & 0.17-0.30 & 80 $\pm$ 5 & 0.2 $\pm$ 0.1 & 20 $\pm$ 5 \\
$[\mathrm{GeV}^{2}]$  & 0.30-0.50 & 74 $\pm$ 4 & 0.3 $\pm$ 0.1 & 26 $\pm$ 5 \\
             & 0.50-1.20 & 72 $\pm$ 4 & 0.3 $\pm$ 0.2 & 28 $\pm$ 5 \\\hline\hline
             & 0.03-0.07 & 90 $\pm$ 4 & 0.2 $\pm$ 0.3 &  10 $\pm$ 4 \\
$x_{B}$     & 0.07-0.10 & 77 $\pm$ 5 & 0.3 $\pm$ 0.1 & 23 $\pm$ 6 \\
             & 0.10-0.15 & 74 $\pm$ 4 & 0.2 $\pm$ 0.1 & 26 $\pm$ 6 \\
             & 0.15-0.35 & 64 $\pm$ 4 & 0.2 $\pm$ 0.1 & 36 $\pm$ 5 \\\hline\hline
             & 1.00-1.50 & 82 $\pm$ 6 & 0.2 $\pm$ 0.1 & 18 $\pm$ 7 \\
$Q^{2}$      & 1.50-2.30 & 74 $\pm$ 4 & 0.2 $\pm$ 0.1 & 26 $\pm$ 6 \\
$[\mathrm{GeV}^{2}]$  & 2.30-3.50 & 80 $\pm$ 4 & 0.3 $\pm$ 0.1 & 20 $\pm$ 5 \\
             & 3.50-10.0 & 75 $\pm$ 3 & 0.2 $\pm$ 0.1 & 25 $\pm$ 3 \\\hline
 \end{tabular}
\end{center}
\caption{
\label{tab:frac2} Monte-Carlo-estimated fractional contributions
to the measured yields by $\,$
$ep\rightarrow e\gamma\pi^+n$, $ep\rightarrow e\gamma p$,
and SIDIS reactions in the selected sample of $ep\rightarrow e\gamma\pi^+n$ events.
}
\end{table}

%%%%%%%%%%%%%%%%%%%%%%%%%%%%%%%%%%%%%%%%%%%%%%%%%%%%%%%%%%%%%%%%%%%%%%%%%%%%%%%%%%%%%%%%%%%%%%%%%%%
%%%%%%%%%%%%%%%%%%%%%%%%%%%%%%%%%%%%%%%%%%%%%%%%%%%%%%%%%%%%%%%%%%%%%%%%%%%%%%%%%%%%%%%%%%%%%%%%%%%
%%%%%%%%%%%%%%%%%%%%%%%%%%%%%%%%%%%%%%%%%%%%%%%%%%%%%%%%%%%%%%%%%%%%%%%%%%%%%%%%%%%%%%%%%%%%%%%%%%%

\section{Extraction of asymmetry amplitudes}
\label{sec:formalism}
Fourier amplitudes of the single-charge beam-helicity asymmetry $\mathcal{A}_{\mathrm{LU}}(\phi;e_\ell)$
are extracted in a manner similar to that applied in ref.~\cite{PublicationDraft68}.
The extraction is based on an extended maximum-likelihood fit
\cite{MML}, unbinned in the azimuthal angle $\phi$.

The distribution of the expectation value of the yield for scattering of a
longitudinally polarized positron beam with polarization $P_\ell$ from an unpolarized hydrogen target is given by

\begin{equation}
\langle\mathcal{N}\rangle(\phi; e_\ell, P_\ell)=\mathcal{L}(e_\ell, P_\ell)\eta(\phi)\sigma_{\mathrm{UU}}(\phi)\left[1+P_\ell\mathcal{A}_{\mathrm{LU}}(\phi; e_\ell)\right],
\label{eq:yield}
\end{equation}

\noindent where $\mathcal{L}$ denotes the integrated luminosity determined by counting
inclusive DIS events and $\eta$ the detection efficiency. 
The asymmetry $\mathcal{A}_{\mathrm{LU}}(\phi;e_\ell)$ is expanded in terms of 
harmonics in $\phi$ in order to extract azimuthal asymmetry amplitudes: 

\begin{equation}
\mathcal{A}_{\mathrm{LU}}(\phi;e_\ell)\simeq A_{\mathrm{LU}}^{\sin\phi}\sin\phi+A_{\mathrm{LU}}^{\sin(2\phi)}\sin(2\phi),
\label{eq:aexpansion}
\end{equation}

\noindent 
where the approximation is due to the truncation of the infinite Fourier series.

As a test of the normalization of the fit, the maximum-likelihood fit is repeated
including the term $A_{\mathrm{LU}}^{\cos(0\phi)}$.
This term is found to be compatible with zero within statistical uncertainties and
to have negligible impact on the resulting asymmetry amplitudes.

%%%%%%%%%%%%%%%%%%%%%%%%%%%%%%%%%%%%%%%%%%%%%%%%%%%%%%%%%%%%%%%%%%%%%%%%%%%%%%%%%%%%%%%%%%%%%%%%%%%
%%%%%%%%%%%%%%%%%%%%%%%%%%%%%%%%%%%%%%%%%%%%%%%%%%%%%%%%%%%%%%%%%%%%%%%%%%%%%%%%%%%%%%%%%%%%%%%%%%%
%%%%%%%%%%%%%%%%%%%%%%%%%%%%%%%%%%%%%%%%%%%%%%%%%%%%%%%%%%%%%%%%%%%%%%%%%%%%%%%%%%%%%%%%%%%%%%%%%%%

\section{Background corrections and systematic uncertainties}
\label{sec:background}

The Monte Carlo simulation shows that the selected samples of associated events contain
contributions from two different sources of background.
The most significant contribution originates from SIDIS production
of neutral pions from the fragmenting struck quark, $ep\rightarrow e\pi^{0}X$, 
with the hadronic system $X$ containing a pion or proton in the recoil detector.
According to the Monte Carlo simulation, its contribution varies
from 8$\%$ to 18$\%$ in the case of the channel 
$ep\rightarrow e\gamma \pi^0 p$ and from 10$\%$ to 36$\%$ in the case of the channel
$ep\rightarrow e\gamma \pi^+ n$, depending on the kinematic bin
(see tables~\ref{tab:frac1} and \ref{tab:frac2}).
The second source of background is the $ep\rightarrow e\gamma p$ reaction,
contributing from 1$\%$ to 14$\%$ for the channel 
$ep\rightarrow e\gamma \pi^0 p$ and negligibly for the channel $ep\rightarrow e\gamma \pi^+ n$.

The asymmetry amplitudes $A_{\mathrm{SIDIS}}$ are extracted from experimental data
using information from only the forward spectrometer.
This approach is based on the assumption that the asymmetry for SIDIS $\pi^0$
production is little affected by the requirement of the detection in the recoil detector
of either a proton or a $\pi^+$ satisfying the kinematic fit for the associated reaction.
Monte Carlo studies showed~\cite{yeth} that the asymmetry extracted for SIDIS $\pi^0$
production is insensitive to event selection using one or two photons.
Thus, in order to estimate the asymmetry of semi-inclusive $\pi^{0}$ background from data,
a ``two-photon analysis'' is performed.
Instead of requiring one trackless cluster in the calorimeter, two trackless clusters
are selected with the energy deposition in the preshower detector larger than 1 MeV.
In addition, the energy of the leading photon is required to be larger than 8 GeV
and the energy of the non-leading one to be above 1 GeV.
The beam-helicity asymmetry amplitudes are extracted with the same maximum-likelihood
fit method as for the associated sample and are found to be consistent with zero.
These asymmetry amplitudes are used to correct for the contribution from the SIDIS reaction in both
the $ep\rightarrow e\gamma \pi^0 p$ and $ep\rightarrow e\gamma \pi^+ n$ channels. 
In order to correct for the small contribution from $ep\rightarrow e\gamma p$, its 
beam-helicity asymmetry amplitude $A_{e\gamma p}$ measured with kinematically complete event reconstruction~\cite{PublicationDraft92} is used. The slightly different kinematic constraints
applied there are not expected to significantly affect this small correction.

The measured asymmetry amplitudes $A_{\mathrm{meas.}}$ are corrected for the above mentioned sources of 
background according to:

\begin{eqnarray}
A_{\mathrm{corr.}}=\frac{A_{\mathrm{meas.}}-f_{e\gamma p}A_{e\gamma p}-f_{\mathrm{SIDIS}}A_{\mathrm{SIDIS}}}{1-f_{e\gamma p}-f_{\mathrm{SIDIS}}} ,
\label{eq:bgcor}
\end{eqnarray}

\noindent
where $f_{e\gamma p}$ and $f_{\mathrm{SIDIS}}$ are the simulated fractional contributions
to the yield from the $ep\rightarrow e\gamma p$
and SIDIS reactions and $A_{e\gamma p}$ and $A_{\mathrm{SIDIS}}$ the corresponding measured
asymmetry amplitudes.
The magnitude of the difference between corrected and measured amplitudes
is assigned as systematic uncertainty (see tables~\ref{tab:syst1} and \ref{tab:syst2}).
This approach takes into account the observed differences between data
and Monte Carlo simulations
presented in figures~\ref{fig:asso_chi_6_8} and \ref{fig:asso_txbq2_6_8}.

In addition to systematic uncertainties due to the background correction described above,
the remaining sources of systematic uncertainties on the extracted asymmetry amplitudes
arise from the spectrometer and recoil-detector acceptance, smearing, and finite bin width.
In order to estimate the combined contribution to the systematic uncertainty from these
three sources, the so-called ``all-in-one'' method is used, which was first employed
in the analysis described in ref.~\cite{PublicationDraft68} and was also used by the latest DVCS
analyses \cite{PublicationDraft69, PublicationDraft90, PublicationDraft92}.
Due to the lack of knowledge about the associated DVCS process, there is no applicable
(GPD) model for use in the Monte Carlo generator, leaving only the BH process with
no interference to produce a beam-helicity asymmetry.
For an estimate of the above mentioned systematic effects, an artificial
$t$-dependent asymmetry of
the expected asymptotic form $A(-t)=C\sqrt{-t}\sin(\phi)+0\sin(2\phi)$ is implemented
for the associated BH process.
The following values of the constant parameter $C$ are applied on generator level:
$C=\{-0.4, -0.2, 0.1, 0.3, 0.5\}$.
(None of these values are conclusively excluded by the experimental data).
The Monte Carlo samples are generated for each beam-polarization state separately, 
passed through a detailed GEANT \cite{GEANT} simulation of the {\sc Hermes} forward spectrometer
and recoil detector, and reconstructed with the same reconstruction and analysis algorithms
as for real data.
After selection of the associated Monte Carlo sample, the maximum-likelihood fit
is performed to extract asymmetry amplitudes in each kinematic bin, referred to
as reconstructed asymmetry amplitudes.
The estimate of the systematic uncertainty due to acceptance, smearing, and finite bin width
is obtained as the difference between the reconstructed Monte Carlo 
asymmetry amplitudes and
those calculated at the reconstructed mean values
of $-t$, $x_B$, and $Q^2$ in each kinematic bin.
The procedure is repeated for each implemented asymmetry separately for both associated channels.
The all-in-one systematic uncertainties are taken as the root mean square of the differences between
reconstructed and calculated asymmetry amplitudes for all parameter values of the
implemented asymmetry, and are presented in tables~\ref{tab:syst1} and \ref{tab:syst2}.

\begin{table}[h!]
\footnotesize 
\begin{center}
\begin{tabular}{|c|c|c|c|c|c|c|c|}
\hline
\multicolumn{2}{|c|}{} & \multicolumn{3}{|c|}{$\delta_{syst} A_{LU}^{\sin\phi}$ ($ep\rightarrow e\gamma \pi^0 p$)} &
\multicolumn{3}{|c|}{$\delta_{syst} A_{LU}^{\sin(2\phi)}$ ($ep\rightarrow e\gamma \pi^0 p$)} \\ \hline\hline
\multicolumn{2}{|c|}{Kinematic bin}  & Bg. corr. & All-in-one & Total & Bg. corr. & All-in-one & Total   \\\hline\hline
\multicolumn{2}{|c|}{Overall}     & ($-$) 0.013  & 0.008 & 0.016 & ($-$) 0.009  & 0.004 & 0.010   \\\hline\hline
                      & $<$0.17   & ($-$) 0.049  & 0.015 & 0.051 & ($-$) 0.004  & 0.007 & 0.009  \\
 $-t$                 & 0.17-0.30 & ($+$) 0.049  & 0.008 & 0.050 & ($+$) 0.031  & 0.005 & 0.031  \\
$[\mathrm{GeV}^{2}]$  & 0.30-0.50 & ($-$) 0.027  & 0.017 & 0.032 & ($-$) 0.013  & 0.001 & 0.013  \\
                      & 0.50-1.20 & ($-$) 0.043  & 0.011 & 0.044 & ($-$) 0.059  & 0.005 & 0.059  \\\hline\hline
                      & 0.03-0.07 & ($+$) 0.013  & 0.015 & 0.020 & ($-$) 0.026  & 0.009 & 0.027  \\
$x_{B}$              & 0.07-0.10 & ($-$) 0.029  & 0.001 & 0.029 & ($-$) 0.019  & 0.008 & 0.021  \\
                      & 0.10-0.15 & ($-$) 0.013  & 0.006 & 0.014 & ($+$) 0.022  & 0.012 & 0.025  \\
                      & 0.15-0.35 & ($-$) 0.144  & 0.021 & 0.146 & ($+$) 0.097  & 0.013 & 0.098  \\\hline\hline
                      & 1.00-1.50 & ($+$) 0.006  & 0.005 & 0.008 & ($+$) 0.042  & 0.009 & 0.043  \\
$Q^{2}$               & 1.50-2.30 & ($+$) 0.032  & 0.019 & 0.037 & ($-$) 0.059  & 0.009 & 0.060  \\
$[\mathrm{GeV}^{2}]$  & 2.30-3.50 & ($-$) 0.033  & 0.010 & 0.035 & ($+$) 0.017  & 0.009 & 0.019  \\
                      & 3.50-10.0 & ($-$) 0.063  & 0.012 & 0.065 & ($+$) 0.040  & 0.010 & 0.041  \\\hline
 \end{tabular}
\end{center}
\caption{
\label{tab:syst1} Individual contributions to the total systematic uncertainties
from background correction and all-in-one estimates of acceptance, smearing, and finite bin
width effects for the channel $ep\rightarrow e\gamma \pi^0 p$. The sign of the 
background corrections is shown in parentheses.}
\end{table}

\begin{table}[h!]
\footnotesize 
\begin{center}
\begin{tabular}{|c|c|c|c|c|c|c|c|}
\hline
\multicolumn{2}{|c|}{} & \multicolumn{3}{|c|}{$\delta_{syst} A_{LU}^{\sin\phi}$ ($ep\rightarrow e\gamma \pi^+ n$)} &
\multicolumn{3}{|c|}{$\delta_{syst} A_{LU}^{\sin(2\phi)}$ ($ep\rightarrow e\gamma \pi^+ n$)} \\ \hline\hline
\multicolumn{2}{|c|}{Kinematic bin}  & Bg. corr. & All-in-one & Total & Bg. corr. & All-in-one & Total   \\\hline\hline
\multicolumn{2}{|c|}{Overall}     & ($-$) 0.005  & 0.010 & 0.012 & ($-$)  0.027  & 0.011 & 0.029   \\\hline\hline
                      & $<$0.17   & ($-$) 0.010  & 0.001 & 0.010 & ($-$) 0.186  & 0.023 & 0.187  \\
 $-t$                 & 0.17-0.30 & ($+$) 0.044  & 0.016 & 0.047 & ($+$) 0.053  & 0.009 & 0.054  \\
$[\mathrm{GeV}^{2}]$  & 0.30-0.50 &  ($-$) 0.042  & 0.017 & 0.045 & ($+$) 0.002  & 0.004 & 0.005  \\
                      & 0.50-1.20 & ($+$) 0.001  & 0.012 & 0.012 & ($-$) 0.001  & 0.015 & 0.015  \\\hline\hline
                      & 0.03-0.07 & ($-$) 0.003  & 0.010 & 0.011 & ($-$) 0.040  & 0.018 & 0.044  \\
$x_{B}$              & 0.07-0.10 & ($-$) 0.056  & 0.035 & 0.066 & ($-$) 0.023  & 0.012 & 0.025  \\
                      & 0.10-0.15 & ($+$) 0.019  & 0.012 & 0.022 & ($+$) 0.013  & 0.009 & 0.016  \\
                      & 0.15-0.35 & ($+$) 0.025  & 0.022 & 0.034 & ($-$) 0.147  & 0.019 & 0.148  \\\hline\hline
                      & 1.00-1.50 & ($-$) 0.078  & 0.014 & 0.079 & ($-$) 0.027  & 0.020 & 0.034  \\
$Q^{2}$               & 1.50-2.30 & ($-$) 0.014  & 0.004 & 0.015 & ($-$) 0.078  & 0.004 & 0.079  \\
$[\mathrm{GeV}^{2}]$  & 2.30-3.50 & ($-$) 0.009  & 0.024 & 0.026 & ($-$) 0.016  & 0.010 & 0.019  \\
                      & 3.50-10.0 & ($+$) 0.049  & 0.013 & 0.051 & ($+$) 0.011  & 0.023 & 0.025  \\\hline
 \end{tabular}
\end{center}
\caption{
\label{tab:syst2} Individual contributions to the total systematic uncertainties
from background correction and all-in-one estimates of acceptance, smearing, and finite bin
width effects for the channel $ep\rightarrow e\gamma \pi^+ n$. The sign of the 
background corrections is shown in parentheses.}
\end{table}

The impact of trigger inefficiency is studied and found to be negligible.

The resulting systematic uncertainties are calculated as the quadratic sum of 
systematic uncertainties from background correction and all-in-one estimates of acceptance,
smearing, and finite bin width effects.
They are summarized in tables~\ref{tab:syst1} and~\ref{tab:syst2} for each kinematic bin 
for the channels $ep\rightarrow e\gamma \pi^0 p$ and $ep\rightarrow e\gamma \pi^+ n$, respectively.

%%%%%%%%%%%%%%%%%%%%%%%%%%%%%%%%%%%%%%%%%%%%%%%%%%%%%%%%%%%%%%%%%%%%%%%%%%%%%%%%%%%%%%%%%%%%%%%%%%%
%%%%%%%%%%%%%%%%%%%%%%%%%%%%%%%%%%%%%%%%%%%%%%%%%%%%%%%%%%%%%%%%%%%%%%%%%%%%%%%%%%%%%%%%%%%%%%%%%%%
%%%%%%%%%%%%%%%%%%%%%%%%%%%%%%%%%%%%%%%%%%%%%%%%%%%%%%%%%%%%%%%%%%%%%%%%%%%%%%%%%%%%%%%%%%%%%%%%%%%

\section{Results and discussion}
\label{sec:results}

\begin{figure}[t!]
\centerline{
\epsfig{file=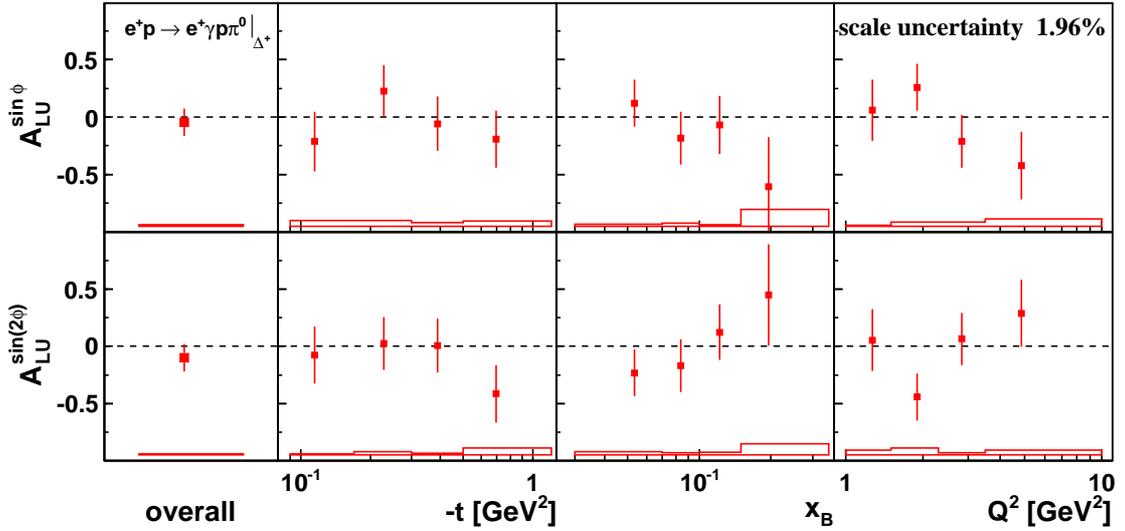, width=1.05\textwidth}
}
\caption{\label{fig:plot-1}  Amplitudes of the single-charge beam-helicity asymmetry extracted in 
the associated channel $ep\rightarrow e\gamma \pi^0 p$ obtained with recoil-proton reconstruction. 
The amplitudes are presented in projections of $-t$, $x_{\textrm{B}}$, and $Q^2$.
The ``overall'' results shown in the very left panel are extracted in a single kinematic bin
covering the entire kinematic acceptance.
Statistical (systematic) uncertainties are represented by error bars (bands).
A separate scale uncertainty arising from the measurement of the beam polarization
amounts to 1.96\%.}
\end{figure}

\begin{figure}[t!]
\centerline{
\epsfig{file=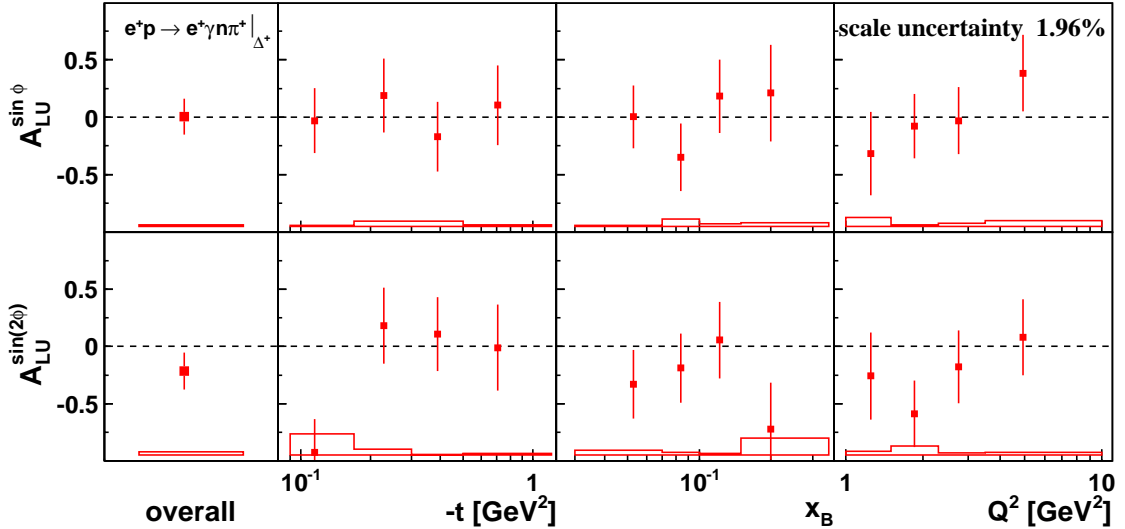, width=1.05\textwidth}
}
\caption{\label{fig:plot-2}  Amplitudes of the single-charge beam-helicity asymmetry extracted in 
the associated channel $ep\rightarrow e\gamma \pi^+ n$ obtained with recoil-pion reconstruction. 
Otherwise as for figure~\ref{fig:plot-1}.}
\end{figure}

Results on asymmetry amplitudes corrected for background contributions are presented
in figures~\ref{fig:plot-1} and \ref{fig:plot-2}, and in tables~\ref{tab:s1} and \ref{tab:s2}.
Each of the asymmetry amplitudes is shown extracted in one bin covering
the entire kinematic region (``overall'') and also projected against $-t$, $x_B$, and $Q^2$.
The beam-helicity asymmetry amplitudes are subject to an additional scale uncertainty of 1.96$\%$ due to the measurement of the beam polarization.
All asymmetry amplitudes are found to be consistent with zero within large
experimental uncertainties.

\begin{table}[h!]
%\scriptsize 
\footnotesize 
\begin{center}
\begin{tabular}{|r|c|c|c|c|c|r|r|}
\hline
\multicolumn{2}{|c|}{Kinematic bin}&Number
&$\langle -t \rangle$ &$\langle x_B \rangle$ &$\langle Q^2
\rangle$ &$A_{LU}^{\sin \, \phi}$ \, \, \, \, \, \, 
&$A_{LU}^{\sin \, (2\phi)}$\, \, \, \, \, \\
\multicolumn{2}{|c|}{} & of events & [GeV$^2$] & &[GeV$^2$] & $\pm \rm \delta_{stat} \pm \rm \delta_{syst}$ \, \, & $\pm \rm \delta_{stat} \pm \rm \delta_{syst}$ \, \, \\
\hline\hline
\multicolumn{2}{|c|}{Overall} & 1185 & 0.35 & 0.10 & 2.54 & $-0.05\pm0.12
\pm0.02$ & $-0.10\pm0.12\pm0.01$ \\
\hline\hline
\multirow{4}{*}{\rotatebox{90}{\mbox{$-t$[GeV$^2$]}}}
&0.00-0.17 & 305 & 0.12& 0.07 & 1.84 & $-0.21\pm0.26\pm0.05$
& $-0.08\pm0.25\pm0.01$ \\
&0.17-0.30 & 303 & 0.23 & 0.09 & 2.38 & $ 0.23\pm0.22\pm0.05$
& $0.03\pm0.23\pm0.03$ \\
&0.30-0.50 & 304 & 0.39 & 0.11 & 2.74 & $-0.06\pm0.24\pm0.03$
& $0.01\pm0.25\pm0.01$ \\
&0.50-1.20 & 273 & 0.69 & 0.12 & 3.27 & $-0.49\pm0.30\pm0.04$
& $-0.55\pm0.33\pm0.06$ \\
\hline\hline
\multirow{4}{*}{\rotatebox{90}{\mbox{$x_B$}}}
&0.03-0.07 & 417 & 0.30 & 0.05 & 1.49 & $0.12\pm0.20\pm0.02$
& $-0.23\pm0.21\pm0.03$ \\
&0.07-0.10 & 318 & 0.28 & 0.08 & 2.16 & $-0.18\pm0.23\pm0.03$
& $-0.17\pm0.23\pm0.02$ \\
&0.10-0.15 & 290 & 0.39 & 0.12 & 3.11 & $-0.07\pm0.25\pm0.01$
& $0.12\pm0.24\pm0.03$ \\
&0.15-0.35 & 160 & 0.54 & 0.20 & 4.99 & $-0.61\pm0.43\pm0.15$
& $0.45\pm0.44\pm0.10$ \\
\hline\hline
\multirow{4}{*}{\rotatebox{90}{\mbox{$Q^{2}$[GeV$^2$]}}}
&1.00-1.50 & 294 & 0.26 & 0.05 & 1.27 & $0.06\pm0.27\pm0.01$
& $0.05\pm0.27\pm0.04$ \\
&1.50-2.30 & 364 & 0.31 & 0.08 & 1.89 & $0.26\pm0.20\pm0.04$
& $-0.44\pm0.20\pm0.06$ \\
&2.30-3.50 & 304 & 0.38 & 0.11 & 2.84 & $-0.21\pm0.23\pm0.04$
& $0.07\pm0.23\pm0.02$ \\
&3.50-10.0 & 223 & 0.49 & 0.17 & 4.85 & $-0.42\pm0.30\pm0.07$
& $0.29\pm0.29\pm0.04$ \\
\hline
\end{tabular}
\caption{\label{tab:s1} Results on amplitudes
extracted in the associated channel $ep\rightarrow e\gamma \pi^0 p$.}
\end{center}
\end{table}

\begin{table}[h!]
%\scriptsize 
\footnotesize 
\begin{center}
\begin{tabular}{|r|c|c|c|c|c|r|r|}
\hline
\multicolumn{2}{|c|}{Kinematic bin}&Number
&$\langle -t \rangle$ &$\langle x_B \rangle$ &$\langle Q^2
\rangle$ &$A_{LU}^{\sin \, \phi}$ \, \, \, \, \, \, 
&$A_{LU}^{\sin \, (2\phi)}$\, \, \, \, \, \\
\multicolumn{2}{|c|}{} & of events & [GeV$^2$] & &[GeV$^2$] & $\pm \rm \delta_{stat} \pm \rm \delta_{syst}$ \, \, & $\pm \rm \delta_{stat} \pm \rm \delta_{syst}$ \, \, \\
\hline\hline
\multicolumn{2}{|c|}{Overall} & 653 & 0.32 & 0.10 & 2.57 & $0.01\pm0.15
\pm0.01$ & $-0.21\pm0.16\pm0.03$ \\
\hline\hline 
\multirow{4}{*}{\rotatebox{90}{\mbox{$-t$[GeV$^2$]}}}
&0.00-0.17 & 218 &  0.12 & 0.08 & 1.90 & $-0.03\pm0.28\pm0.01$
& $-0.93\pm0.29\pm0.19$ \\
&0.17-0.30 & 154 & 0.23 & 0.10 & 2.49 & $0.19\pm0.32\pm0.05$
& $0.18\pm0.33\pm0.05$ \\
&0.30-0.50 & 156 & 0.39 & 0.11 & 2.88 & $-0.17\pm0.30\pm0.05$
& $0.11\pm0.32\pm0.01$ \\
&0.50-1.20 & 125 & 0.71 & 0.12 & 3.47 & $0.11\pm0.35\pm0.01$
& $-0.01\pm0.38\pm0.02$ \\
\hline\hline
\multirow{4}{*}{\rotatebox{90}{\mbox{$x_B$}}}
&0.03-0.07 & 228 & 0.28 & 0.05 & 1.48 & $0.00\pm0.27\pm0.01$
& $-0.33\pm0.30\pm0.04$ \\
&0.07-0.10 & 183 & 0.28 & 0.08 & 2.20 & $-0.35\pm0.29\pm0.07$
& $-0.19\pm0.30\pm0.03$ \\
&0.10-0.15 & 156 & 0.34 & 0.12 & 3.13 & $0.18\pm0.32\pm0.02$
& $0.06\pm0.33\pm0.02$ \\
&0.15-0.35 & 86 & 0.49 & 0.20 & 5.26 & $0.21\pm0.42\pm0.03$
& $-0.72\pm0.41\pm0.15$ \\
\hline\hline
\multirow{4}{*}{\rotatebox{90}{\mbox{$Q^{2}$[GeV$^2$]}}}
&1.00-1.50 & 158 & 0.24 & 0.05 & 1.25 & $-0.32\pm0.36\pm0.08$
& $-0.26\pm0.38\pm0.03$ \\
&1.50-2.30 & 189 & 0.26 & 0.08 & 1.85 & $-0.08\pm0.28\pm0.02$
& $-0.59\pm0.29\pm0.08$ \\
&2.30-3.50 & 173 & 0.35 & 0.10 & 2.76 & $-0.03\pm0.29\pm0.03$
& $-0.18\pm0.32\pm0.02$ \\
&3.50-10.0& 133 & 0.47 & 0.17 & 4.93 & $0.38\pm0.33\pm0.05$
& $0.08\pm0.33\pm0.03$ \\
\hline
\end{tabular}
\caption{\label{tab:s2} Results on amplitudes
extracted in the associated channel $ep\rightarrow e\gamma \pi^+ n$.}
\end{center}
\end{table}

The model of ref.~\cite{Asso03} described in
section~\ref{sec:Introduction}, employing the VGG model \cite{Vdh99, GPV01}
for the nucleon GPDs, predicts
the $\sin \phi$ asymmetry amplitudes to be about $-0.15$ in the case of the
$ep\rightarrow e\gamma \pi^0 p$ channel and about $-0.10$ in the case of the
$ep\rightarrow e\gamma \pi^+ n$ channel.\footnote{In ref. \cite{Asso03},
a different convention for the $\phi$ angle definition was used leading to the
opposite sign of asymmetry amplitudes.}
The presented experimental results do not exclude this model.

Recently, {\sc Hermes} published results on the single-charge beam-helicity asymmetry
arising from DVCS with kinematically complete event reconstruction \cite{PublicationDraft92}.
The main result of this publication was that after removal of associated background from the
data sample the magnitude of the leading asymmetry amplitude increased.
This increase is consistent with the small magnitude of the asymmetries
in the two associated channels obtained in this analysis.
Effectively, the background from the associated reaction acts as a dilution in the
beam-helicity asymmetries
measured previously by {\sc Hermes} using the missing-mass technique
\cite{PublicationDraft69, PublicationDraft90}.

%%%%%%%%%%%%%%%%%%%%%%%%%%%%%%%%%%%%%%%%%%%%%%%%%%%%%%%%%%%%%%%%%%%%%%%%%%%%%%%%%%%%%%%%%%%%%%%%%%%
%%%%%%%%%%%%%%%%%%%%%%%%%%%%%%%%%%%%%%%%%%%%%%%%%%%%%%%%%%%%%%%%%%%%%%%%%%%%%%%%%%%%%%%%%%%%%%%%%%%
%%%%%%%%%%%%%%%%%%%%%%%%%%%%%%%%%%%%%%%%%%%%%%%%%%%%%%%%%%%%%%%%%%%%%%%%%%%%%%%%%%%%%%%%%%%%%%%%%%%

\section{Summary}
\label{sec:summary}

Amplitudes of the beam-helicity asymmetry are measured at {\sc Hermes}
in exclusive associated production of real photons, $ep\rightarrow e\gamma \pi N$,
by longitudinally polarized positrons incident on an unpolarized hydrogen target.  
The selected $ep\rightarrow e\gamma \pi^0 p$ ($ep\rightarrow e\gamma \pi^+ n$)
event sample is estimated to contain on average 11\% (23\%) contribution
from SIDIS production, which is corrected for in the analysis.
Corrections for the small contributions from $ep\rightarrow e\gamma p$
are applied using asymmetry amplitudes obtained previously by {\sc Hermes}.
All asymmetry amplitudes are found to be consistent with zero within experimental uncertainties
that are at best $\pm0.12$ in the full acceptance. The only available theoretical
estimates \cite{Asso03} for the asymmetry amplitudes are consistent with the measurements.
This finding may offer support for the model of transition GPDs
in terms of nucleon GPDs, based on the soft-pion technique and the large $N_c$ limit.

%%%%%%%%%%%%%%%%%%%%%%%%%%%%%%%%%%%%%%%%%%%%%%%%%%%%%%%%%%%%%%%%%%%%%
\acknowledgments

We gratefully acknowledge the {\sc Desy} management for its support and the staff
at {\sc Desy} and the collaborating institutions for their significant effort.
This work was supported by 
the Ministry of Economy and the Ministry of Education and Science of Armenia;
the FWO-Flanders and IWT, Belgium;
the Natural Sciences and Engineering Research Council of Canada;
the National Natural Science Foundation of China;
the Alexander von Humboldt Stiftung,
the German Bundesministerium f\"ur Bildung und Forschung (BMBF), and
the Deutsche Forschungsgemeinschaft (DFG);
the Italian Istituto Nazionale di Fisica Nucleare (INFN);
the MEXT, JSPS, and G-COE of Japan;
the Dutch Foundation for Fundamenteel Onderzoek der Materie (FOM);
the Russian Academy of Science and the Russian Federal Agency for 
Science and Innovations;
the Basque Foundation for Science (IKERBASQUE) and the UPV/EHU under program UFI 11/55;
the U.K.~Engineering and Physical Sciences Research Council, 
the Science and Technology Facilities Council,
and the Scottish Universities Physics Alliance;
the U.S.~Department of Energy (DOE) and the National Science Foundation (NSF);
as well as the European Community Research Infrastructure Integrating Activity
under the FP7 "Study of strongly interacting matter (HadronPhysics3, Grant
Agreement number 283286)".

%%%%%%%%%%%%%%%%%%%%%%%%%%%%%%%%%%%%%%%%%%%%%%%%%%%%%%%%%%%%%%%%%%

%\clearpage

\bibliographystyle{natbib}

%%%%%%%%%%%%%%%
\end{document}